# Progress on Integrating Quantum Communications in Optical Systems Testbeds

Jerry Horgan
Walton Institute
South East Technological University
Waterford, Ireland
jerry.horgan@waltoninstitute.ie

Dmitrii Briantcev
CONNECT Centre
Trinity College Dublin
Dublin, Ireland
briantcd@tcd.ie

Aleksandra Kaszubowska-Anandarajah
Department of Electronic & Electrical Engineering
Trinity College Dublin
Dublin, Ireland
anandara@tcd.ie

Marco Ruffini
School of Computer Science and Statistics
Trinity College Dublin
Dublin, Ireland
ruffinm@tcd.ie

Dan Kilper
CONNECT Centre
Trinity College Dublin
Dublin, Ireland
dan.kilper@tcd.ie

*Abstract*—Experimental methods are being developed to enable quantum communication systems research in testbeds. We describe testbed architectures for emerging quantum technologies and how they can integrate with existing fibre optical testbeds, specifically OpenIreland.

*Keywords—quantum communications, optical systems testbeds, quantum key distribution, entanglement distribution*

## I. INTRODUCTION

Testbeds have played a central role in the research and development of fibre optical transmission systems, which are the foundation for today's global Internet. Recent progress in quantum computing is fuelling the development of quantum communication systems to provide quantum connectivity between quantum computers and devices to form a quantum Internet. Quantum security technologies such as quantum key distribution (QKD) are already being used in commercial networks. While QKD systems are a starting point for quantum communication networks, distributed quantum computing will require the distribution of quantum entanglement through optical networks, which is not required for QKD. The interfaces between quantum computing elements, the fibre network, and even the qubit or entanglement transmission format are still open questions. Many network elements, such as quantum repeaters, are still just theorized or have primarily been studied in isolated lab experiments, such as quantum memories. Similar to early fibre optical internet testbeds, there is a potential for quantum system testbeds to serve as an essential tool in accelerating the development of component technologies and growing nascent research on related systems and networks. Connecting a quantum communications testbed with an existing fibre optical communication system testbed, such as OpenIreland, would provide experimentation on quantum communication systems together with classical optical systems either coexisting or providing control and management functions.

## II. OPTICAL NETWORK TESTBEDS

### A. Early Fibre Internet

As many key quantum technologies still need to be developed and system protocols and architectures have only been proposed, it may seem premature to begin the development of quantum communication system testbeds. Historically, the early fibre optical Internet technology development provides evidence that this is not the case. The Gigabit Testbeds in the 1980s were a national-scale testbed effort composed of federated testbeds at different locations across the US. At the time, a viable fibre amplification technology did not exist, and there needed to be a consensus on the key transmission technologies, protocols, and architectures. Most of the component technologies were emerging from physics research labs. Different such technologies were explored through this federated set of testbeds and provided insights to help move towards viable large-scale systems [1]. There were, in fact, numerous testbeds in the 1980s and 1990s, leading up to the fibre boom in 1998 in which optical amplification and wavelength division multiplexing matured, and the community converged on the optical transmission architectures that would evolve into the architectures widely used today.

Another notable testbed is the MONET project, which was key to developing wavelength division multiplexing (WDM) technologies [2]. Many in the field point to this project as being the precursor to WDM system networking. Although MONET was very influential in progressing WDM networking technologies, the technologies investigated in MONET were far before their commercial introduction.

TABLE I.        1999 MONET Testbed Technology Evolution

| No | Description | Year Adoption |
|---|---|---|
| 1. | Acousto-optic tunable filters | N/A |
| 2. | Coherent Transceivers | 2010 |
| 3. | Wavelength crossconnects & ADMs | 2004 |
| 4. | Tunable Transceivers | 2004 |
| 5. | Mesh Networks | 2006 |
| 6. | Space division multiplexing (parallel fibres) | TBD |

Table one lists key technologies investigated in MONET and the approximate year they experienced large-scale commercial deployment. Some technologies, such as acousto-optic tunable filters, have never been commercially deployed at scale in fibre-optic transmission systems. High-speed DSP improvements were needed to use coherent transceivers, which required another decade of development. Wavelength cross-connects, add-drop multiplexers, and tunable lasers are all key elements for optical networking and were introduced when the wavelength selective switch was developed into a reconfigurable optical add-drop multiplexer (ROADM). Mesh

networks soon followed. Spatial division multiplexing in the form of components specifically designed for parallel fibre systems is rapidly approaching commercialization as traffic demand has been catching up with fibre transmission band capacity limits.

Thus, a research testbed can incorporate many technologies that might be decades from commercialization while still influencing near-term systems development. In this way, testbeds can establish the relative readiness of the component technologies. This enables a convergence toward viable technologies while also identifying the research needs of technologies that are not yet ready but might still find application in the future.

### B. City-Scale Testbeds

Recent optical transmission testbeds have evolved into city-scale platforms that enable a combination of lab and deployed fibre plants for experimentation. Unlike lab-only recirculating loop transmission testbeds, these city-scale testbeds provide two critical capabilities for transmission research today: 1) data collection on deployed fibre systems for AI applications and 2) large-scale networking for network control experimentation. Examples of such testbeds include Bristol is Open [3], COSMOS [4], and OpenIreland [5].

In these city-scale testbeds, large fibre (space) switch fabrics are a key element to facilitate experimentation. These switches serve a dual purpose: they can be used as a testbed control element to reconfigure component and fibre connections to realize a wide range of experimental configurations and network topologies. At the same time, they can be used as a networking element within the respective architectures under investigation [6].

Another important element of these testbeds is software-defined networking (SDN) controls and emulation. Here again, a dual-use approach is taken in which the SDN controller can be used to control and configure experiments, and it can be used as part of the architecture under experimentation [7]. The SDN platforms can be made compatible with Mininet or Mininet Optical in order to enable virtual system emulation or digital twin capabilities [8].

### III. INTEGRATING QUANTUM CAPABILITIES

The use of a space switch on the line side, connecting the transmission fibre is advantageous for integrating quantum elements into the city-scale testbed platforms. The line side quantum signals can either be switched directly to the quantum transceivers or to a filtering element that can separate the quantum signals from other signals, bypassing amplifiers and other incompatible components. This allows for experimental flexibility, but at the cost of additional loss. Fortunately, fibre space switches typically have very low loss, below 3 dB port to port. For the case in which the testbed fibre can support either quantum or classical signals, elements are needed to allow the quantum signals to bypass optical amplifiers and to strongly filter out any light such as amplifier ASE noise from the wavelengths carrying the quantum signals. If the quantum signals, typically at single photon power levels, propagate in the same fibre with classical signals then a number of linear and non-linear impairments need to be taken into account [9, 10]. These include linear crosstalk due to imperfect filtering or blocking of light and non-linear fibre effects such as Raman scattering and four-wave mixing [11].

### A. Coexistance Technologies

Several strategies for successful coexisting of classical and quantum signals have been studied. These include band splitting (O-band quantum signals/C-band classical signals) [12, 13], spectral guard bands (>1600GHz) [14-16], classical signal power limiting, e.g. to less than -15dBm, [14-18], or physical isolation through novel fibre technology such as multi-core fibre [18-20], few-mode fibre [21,22], or hollow-core fibre [23] to reduce or eliminate the non-linear effects in particular.

Multi-core and hollow core fibres are particularly effective at supporting the coexistence of quantum and classical signals. Experiments have shown coexistence of QKD and high throughput classical channels, achieving 9.6 Tbps [19] and 11.2 Tbps [20] using multi-core fibre, and 1.6 Tbps [23] using hollow-core fibre. Hollow core works by suppressing the nonlinear fibre effects due to the extreme reduction of core material, whereas multi-core fibres physically isolate the quantum and classical signals into separate fibre cores. Due to the limited availability of these fibres, experiments have been limited to 1-2km of fibre. The hollow-core fibre setup provided very high isolation between the quantum and classical channels (>110dB), while the multi-core fibre setups used reduced classical launch powers to limit cross talk.

Multimode fibre also allows for spatial isolation. In a 37-core fibre experiment quantum and classical signal were co-propagated over 8km [18]. The mode selectivity of the cascaded beamsplitter demultiplexer achieved a secret key rate of 105.7 Mbps and a classical throughput of 370Gbps. However, the classical signal launch power was limited to -34 dBm and SNSPDs were used at the detector.

### B. QKD Testbeds

The first known field deployed QKD experiment was by DARPA in Boston in 2003 [24] and a series of field trials followed [25]. A major step forward occurred with the deployment of the Beijing-Shanghai backbone testbed between 2013 and 2016 of over 2,000km from Beijing to Shanghai [26] using 32 nodes, and the satellite relayed intercontinental link of over 7,600 km from China to Europe [27]. This testbed includes both terrestrial fibre links and satellite links.

A long-term (580 days) 3-node metropolitan QKD network was deployed in Cambridge, UK [28]. The network was over 25km in length, produced 120 Tb of secure key over its lifetime, and operated 2 x 100 Gbps classical channels, which were encrypted using the secret keys produced.

The Madrid Quantum Communication Infrastructure (MadQCI) testbed is a multi-ring QKD network [29] providing a multi-vendor, multi-operator, and multi-user environment. The outer research ring is 250 km in length, while the inner core Telefonica production network is 28 km. This network mixes continuous variable (CV) QKD and discrete variable (DV) QKD technologies and classical communications on the same fibre, while utilising an SDN based software stack for transparent end-to-end secret key routing, quantum / classical network management, and service integration in a multi-vendor, multi-network topology.

In 2021 researchers from QuTech in the Netherlands and Cisco Systems demonstrated a multi-city Measurement Device Independent (MDI) QKD deployment [30]. The MDI Central Node was deployed in Rijswijk, with Edge Nodes deployed in Delft and Den Haag, 14.7km and 10.2km from the

Central Node respectively. The quantum channel operated in the O-band (1310nm) and the data channel operated in the C-band. Furthermore, Cisco ROADMs were used to demonstrate coexistence with a 10Gbps and 100Mbps classical IP service.

A Quantum Network Service Provider metropolitan testbed was established in Bristol in 2020 [31]. This testbed utilised existing telecommunications fibre across the city of Bristol to provide entanglement distribution across a network of users. The network used a single quantum entanglement source which produced a broad-band spectrum. Using a quantum-ROADM this spectrum was divided into 30x100GHz channels, which were then (dynamically) allocated amongst all end users to provide a fully meshed network. These channels were further split (1:4) to minimise the number of channels required to connect all the network users [32].

### C. Emerging Technologies

Many new technologies and architectures are under investigation today and would benefit from testbed experimentation. Most commercial QKD technology uses weak coherent state light sources. Entangled photon pair sources have unique characteristics and have potential for use in methods to extend the transmission reach. The technology to deploy the Quantum Internet, as envisaged in [33], is still emerging. Unlike classical signals, quantum signals cannot be amplified, therefore quantum repeaters will be necessary for long distance transmission. Most repeater architectures under consideration require memories, or buffers at a minimum, to hold qubits. Most such quantum memories require cryogenic environments often at dilution refrigerator temperatures (milli-kelvin range). Thus, integrating cryogenic environments with testbeds will allow for experiments using these devices. One such testbed is the CQN Boston testbed, which realized the first quantum memory assisted entanglement distribution experiment. A list of potential new technologies for testbed experiments are listed in Table II.

TABLE II. CANDIDATE QUANTUM TESTBED TECHNOLOGIES

| No | Description | Ref |
|---|---|---|
| 1. | Single qubit memories | [34] |
| 2. | Qubit memory arrays | [35] |
| 3. | Quantum signal wavelength/mode convertors | [36] |
| 4. | Bell state measurement (BSM) elements | [37] |
| 5. | Wavelength selective BSM | [38] |
| 6. | Quantum repeaters | [39] |
| 7. | Quantum datagrams | [40] |
| 8. | Time multiplexing | [41] |

Single qubit memories can be used for storing qubits without computation and include e.g. Rubidium gas cell based memories [34]. Qubit memory arrays are arrays of qubits that typically may include computational capabilities [35]. Wavelength and mode convertors are used with most memory technologies in order to shape the mode to match the qubit resonance and convert between the qubit resonant wavelengths and telecom wavelengths (1310 and 1550 nm) [36]. Bell state measurements are used to perform entanglement swapping and to herald entanglement [37]. This can be done using filters or demultiplexers for wavelength selective BSM [38]. Quantum repeaters are memory platforms designed to support entanglement swapping and distillation/purification or quantum error correction coding [39]. Quantum datagrams are formed by inserting a classical signal with header control and management information before one or more quantum signals in time [40]. More general time multiplexing of classical and quantum signals has also been proposed [41].

## IV. OPENIRELAND

OpenIreland is a city-scale testbed that includes 1700 km of spooled lab fibre connected to dark and lit fibre across the metro Dublin area. OpenIreland is designed to connect to other optical fibre and quantum communications testbeds.

The core of OpenIreland is designed around a central fibre space switch or optical cross-connect (OXC) which allows for flexible automated reconfiguration of connections between the many network elements, as shown in Fig. 1.

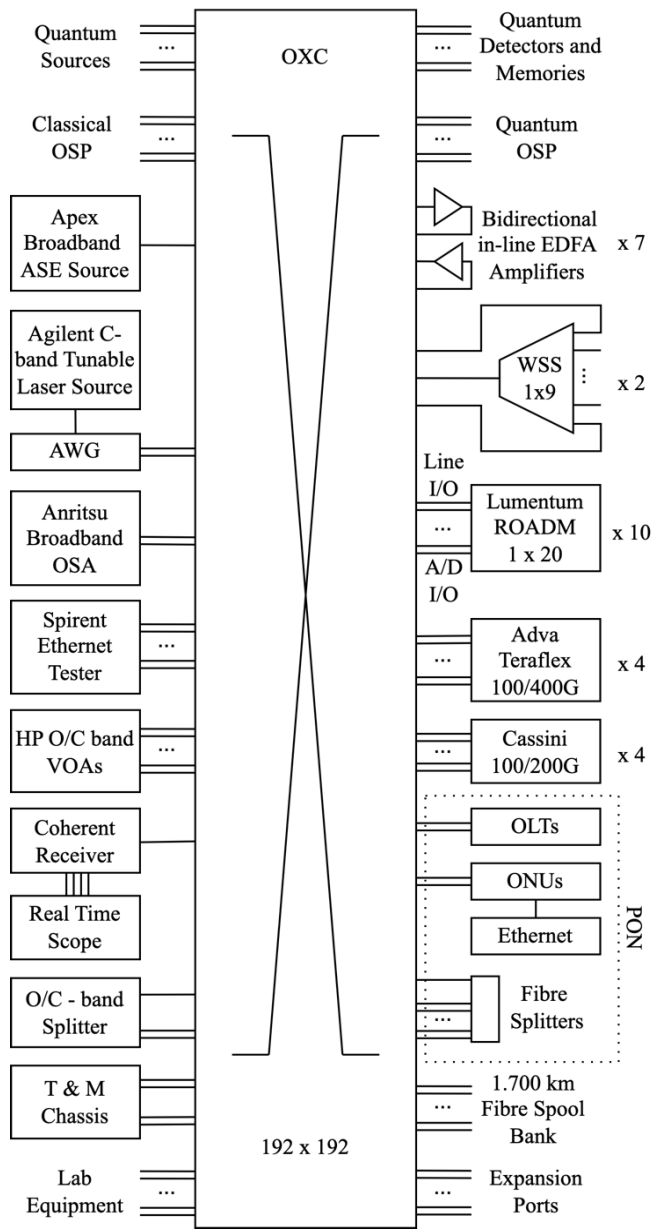

*Figure 1: Central OpenIreland Node*

The OXC has four main categories of connections: 1) line side fibre, 2) transceiver elements, 3) wavelength and transmission management elements, and 4) test and measurement devices. The line side fibre includes both spooled lab fibre and outside plant fibre. Classical transceivers include 100-400G coherent Adva Teraflex and 100/200G coherent Cassini hardware. Classical signals are multiplexed using optically amplified Lumentum ROADMs. A combination of wavelength selective switches, band splitters, and couplers can be used for multiplexing quantum signals together and with classical signals. Optical power is managed through the ROADM/WSS units or dedicated variable optical attenuators. Test equipment includes ports for an OSA, custom coherent receiver and real time scope, and an OTDR port as well as an open port for attaching external test equipment.

Additional quantum expansion ports can be used to attach novel quantum hardware. This might include EPR pair sources, SNSPD detector arrays, MDI-QKD elements and BSM elements. This hardware will be a mix of dedicated equipment installed in the network and custom equipment introduced for specific experiments by users. In addition, cryogenic stations can be connected for housing low temperature devices. These stations would allow for multi-user experiments in which default devices will be available for use, such as custom memory arrays, or users can introduce their own devices for in-network testing.

## ACKNOWLEDGMENT

This work is supported in part by the Science Foundation Ireland (SFI) grant 20/US/3708 and the US National Science Foundation (NSF) grant CNS-2107265, the SFI grant 21/US-C2C/3750, and the SFI CONNECT Centre under grant 13/RC/2077_P2.

## REFERENCES

[1] http://www.cnri.reston.va.us/gigafr/Gigabit_Final_Rpt.pdf
[2] W. T. Anderson and , "The MONET Project-A Final Report," in Optical Fiber Communication Conference, OSA Technical Digest Series (Optica Publishing Group, 2000), paper WJ1.
[3] D. Simeonidou, Bristol is open, in: 5G Radio Technology Seminar. Exploring Technical Challenges in the Emerging 5G Ecosystem, 2015, pp. 1–32. doi:10.1049/ic.2015.0035.
[4] J. Yu et al., "COSMOS: Optical Architecture and Prototyping," 2019 Optical Fiber Communications Conference and Exhibition (OFC), San Diego, CA, USA, 2019, pp. 1-3.
[5] K. Kaeval, F. Slyne, S. Troia, E. Kenny, K. Große, H. Griesser, D.C. Kilper, M. Ruffini, J-J Pedreno-Manresa, S.K. Patri and G. Jervan. Employing Channel Probing to Derive End-of-Life Service Margins for Optical Spectrum Services. OPTICA Journal of Optical Communications and Networking, July 2023.
[6] E. Akinrintoyo, Z. Wang, B. Lantz, T. Chen, and D. Kilper, "(invited)reconfigurable topology testbeds: A new approach to optical system experiments," Optical Fiber Technology, vol. 76, p. 103243, 2023.
[7] T. Chen et al., "A Software-Defined Programmable Testbed for Beyond 5G Optical-Wireless Experimentation at City-Scale," in IEEE Network, vol. 36, no. 2, pp. 90-99, March/April 2022, doi: 10.1109/MNET.006.2100605.
[8] B. Lantz, A. A. Díaz-Montiel, J. Yu, C. Rios, M. Ruffini and D. Kilper, "Demonstration of Software-Defined Packet-Optical Network Emulation with Mininet-Optical and ONOS," 2020 Optical Fiber Communications Conference and Exhibition (OFC), San Diego, CA, USA, 2020, pp. 1-3.
[9] N. Peters, P. Toliver, T. Chapuran, R. Runser, S. McNown, C. Peterson, D. Rosenberg, N. Dallmann, R. Hughes, K. McCabe et al., "Dense wavelength multiplexing of 1550 nm qkd with strong classical channels in reconfigurable networking environments," New Journal of physics, vol. 11, no. 4, p. 045012, 2009.
[10] P. Eraerds, N. Walenta, M. Legré, N. Gisin, and H. Zbinden, "Quantum key distribution and 1 gbps data encryption over a single fibre," New Journal of Physics, vol. 12, no. 6, p. 063027, 2010.
[11] M. Mlejnek, N. A. Kaliteevskiy, and D. A. Nolan, "Reducing spontaneous raman scattering noise in high quantum bit rate qkd systems over optical fiber," arXiv preprint arXiv:1712.05891, 2017.
[12] I. Choi, R. J. Young, and P. D. Townsend, "Quantum information to the home," New Journal of Physics, vol. 13, no. 6, p. 063039, 2011.
[13] J.-Q. Geng, G.-J. Fan-Yuan, S. Wang, Q.-F. Zhang, Y.-Y. Hu, W. Chen, Z.-Q. Yin, D.-Y. He, G.-C. Guo, and Z.-F. Han, "Coexistence of quantum key distribution and optical transport network based on standard single-mode fiber at high launch power," Optics Letters, vol. 46, no. 11, pp. 2573–2576, 2021.
[14] R. Kumar, H. Qin, and R. Alléaume, "Experimental demonstration of the coexistence of continuous-variable quantum key distribution with an intense dwdm classical channel," in CLEO: QELS Fundamental Science. Optica Publishing Group, 2014, pp. FM4A–1.
[15] J. F. Dynes, W. W. Tam, A. Plews, B. Fröhlich, A. W. Sharpe, M. Lucamarini, Z. Yuan, C. Radig, A. Straw, T. Edwards et al., "Ultra-high bandwidth quantum secured data transmission," Scientific reports, vol. 6, no. 1, p. 35149, 2016.
[16] R. Valivarthi, P. Umesh, C. John, K. A. Owen, V. B. Verma, S. W. Nam, D. Oblak, Q. Zhou, and W. Tittel, "Measurement-device-independent quantum key distribution coexisting with classical communication,"Quantum Science and Technology, vol. 4, no. 4, p. 045002, 2019.
[17] K. Patel, J. Dynes, I. Choi, A. Sharpe, A. Dixon, Z. Yuan, R. Penty, and A. Shields, "Coexistence of high-bit-rate quantum key distribution and data on optical fiber," Physical Review X, vol. 2, no. 4, p. 041010, 2012.
[18] D. Bacco, B. Da Lio, D. Cozzolino, F. Da Ros, X. Guo, Y. Ding, Y. Sasaki, K. Aikawa, S. Miki, H. Terai et al., "Boosting the secret key rate in a shared quantum and classical fibre communication system," Communications Physics, vol. 2, no. 1, p. 140, 2019.
[19] E. Hugues-Salas, R. Wang, G. T. Kanellos, R. Nejabati, and D. Simeonidou, "Co-existence of 9.6 tb/s classical channels and a quantum key distribution (qkd) channel over a 7-core multicore optical fibre," in 2018 IEEE British and Irish Conference on Optics and Photonics (BICOP). IEEE, 2018, pp. 1–4.
[20] E. Hugues-Salas, O. Alia, R. Wang, K. Rajkumar, G. T. Kanellos, R. Nejabati, and D. Simeonidou, "11.2 tb/s classical channel coexistence with dv-qkd over a 7-core multicore fiber," Journal of Lightwave Technology, vol. 38, no. 18, pp. 5064–5070, 2020.
[21] B.-X. Wang, Y. Mao, L. Shen, L. Zhang, X.-B. Lan, D. Ge, Y. Gao, J. Li, Y.-L. Tang, S.-B. Tang et al., "Long-distance transmission of quantum key distribution coexisting with classical optical communication over a weakly-coupled few-mode fiber," Optics express, vol. 28, no. 9, pp. 12 558–12 565, 2020.
[22] H. Zhong, S. Zou, D. Huang, and Y. Guo, "Continuous-variable quantum key distribution coexisting with classical signals on few-mode fiber," Optics Express, vol. 29, no. 10, pp. 14 486–14 504, 2021.
[23] O. Alia, R. S. Tessinari, S. Bahrani, T. D. Bradley, H. Sakr, K. Harrington, J. Hayes, Y. Chen, P. Petropoulos, D. Richardson et al., "Dv-qkd coexistence with 1.6 tbps classical channels over hollow core fibre," Journal of Lightwave Technology, vol. 40, no. 16, pp. 5522–5529, 2022.
[24] C. Elliott, A. Colvin, D. Pearson, O. Pikalo, J. Schlafer, and H. Yeh, "Current status of the darpa quantum network," in Quantum Information and computation III, vol. 5815. SPIE, 2005, pp. 138–149.
[25] Y. Cao, Y. Zhao, Q. Wang, J. Zhang, S. X. Ng, and L. Hanzo, "The evolution of quantum key distribution networks: On the road to the qinternet," IEEE Communications Surveys & Tutorials, vol. 24, no. 2, pp. 839–894, 2022.
[26] Q. Zhang, F. Xu, L. Li, N.-L. Liu, and J.-W. Pan, "Quantum information research in china," Quantum Science and Technology, vol. 4, no. 4, p. 040503, 2019.
[27] S.-K. Liao, W.-Q. Cai, J. Handsteiner, B. Liu, J. Yin, L. Zhang, D. Rauch, M. Fink, J.-G. Ren, W.-Y. Liu, Y. Li, Q. Shen, Y. Cao, F.-Z. Li, J.-F. Wang, Y.-M. Huang, L. Deng, T. Xi, L. Ma, T. Hu, L. Li, N.-L. Liu, F. Koidl, P. Wang, Y.-A. Chen, X.-B. Wang, M. Steindorfer, G. Kirchner, C.-Y. Lu, R. Shu, R. Ursin, T. Scheidl, C.-Z. Peng, J.-Y.

Wang, A. Zeilinger, and J.-W. Pan, "Satellite-relayed intercontinental quantum network," Phys. Rev. Lett., vol. 120, p. 030501, Jan 2018.

[28] J. Dynes, A. Wonfor, W.-S. Tam, A. Sharpe, R. Takahashi, M. Lucamarini, A. Plews, Z. Yuan, A. Dixon, J. Cho et al., "Cambridge quantum network," npj Quantum Information, vol. 5, no. 1, p. 101, 2019.

[29] D. Lopez, J. P. Brito, A. Pastor, V. Martín, C. Sánchez, D. Rincon, and V. Lopez, "Madrid quantum communication infrastructure: a testbed for assessing qkd technologies into real production networks," in Optical Fiber Communication Conference. Optica Publishing Group, 2021, pp. Th2A–4.

[30] Berrevoets, R.C., Middelburg, T., Vermeulen, R.F.L. et al. Deployed measurement-device independent quantum key distribution and Bell-state measurements coexisting with standard internet data and networking equipment. Commun Phys 5, 186 (2022).

[31] Siddarth Koduru Joshi et al. ,A trusted node–free eight-user metropolitan quantum communication network. Sci. Adv.6, eaba0959(2020). DOI:10.1126/sciadv.aba0959.

[32] M. J. Clark, O. Alia, R. Wang, S. Bahrani, M. Peranić, D. Aktas, G. T. Kanellos, M. Loncaric, Ž. Samec, A. Radman, M. Stipcevic, R. Nejabati, D. Simeonidou, J. G. Rarity, S. K. Joshi, "Entanglement distribution quantum networking within deployed telecommunications fibre-optic infrastructure," Proc. SPIE 12335, Quantum Technology: Driving Commercialisation of an Enabling Science III, 123350E (11 January 2023);https://doi.org/10.1117/12.2645095

[33] S. Wehner, D. Elkouss, and R. Hanson, "Quantum internet: A vision for the road ahead," Science, vol. 362, no. 6412, Oct. 2018, Art. no. eaam9288.

[34] Y. Wang, A. N. Craddock, R. Sekelsky, M. Flament, and M. Namazi, "Field-deployable quantum memory for quantum networking," Phys. Rev. Appl., vol. 18, p. 044058, Oct 2022. [Online]. Available: https://link.aps.org/doi/10.1103/PhysRevApplied.18.044058

[35] N. H. Wan, T. Lu, K. C. Chen, M. Walsh, M. Trusheim, L. De Santis, E. Bersin, I. Harris, S. Mouradian, I. Christen, E. Bielejec, and D. Englund, "A 128-channel diamond quantum memory array integrated in a microphotonic chip," in Conference on Lasers and Electro-Optics, OSA Technical Digest (Optica Publishing Group, 2020), paper JM3G.4.

[36] Ikuta, R., Kusaka, Y., Kitano, T. et al. Wide-band quantum interface for visible-to-telecommunication wavelength conversion. Nat Commun 2, 537 (2011).

[37] Welte, S., Thomas, P., Hartung, L. et al. A nondestructive Bell-state measurement on two distant atomic qubits. Nat. Photon. 15, 504–509 (2021).

[38] K. C. Chen, P. Dhara, M. Heuck, Y. Lee, W. Dai, S. Guha, and D. Englund, "Zero-added-loss entangled-photon multiplexing for ground- and space-based quantum networks," Phys. Rev. Appl., vol. 19,p. 054029, May 2023.

[39] Wang, H., Trusheim, M.E., Kim, L. et al. Field programmable spin arrays for scalable quantum repeaters. Nat Commun 14, 704 (2023).

[40] S. DiAdamo, B. Qi, G. Miller, R. Kompella, and A. Shabani, "Packet switching in quantum networks: A path to the quantum internet," Phys. Rev. Res., vol. 4, p. 043064, Oct 2022.

[41] Z. Xie, Y. Liu, X. Mo, T. Li, and Z. Li, "Quantum entanglement creation for distant quantum memories via time-bin multiplexing," Phys. Rev. A, vol. 104, p. 062409, Dec 2021.